\documentstyle[11pt,newpasp,twoside,epsf]{article}
\begin{document}
\title{Wide Field Imaging of the Sextans Dwarf Spheroidal Galaxy}
 \author{Elena Pancino}
  \affil{Dipartimento di Astronomia, Universit\`a di Bologna, Via
Ranzani 1, I-40127 Bologna, Italy} 
 \author{Michele Bellazzini and Francesco R. Ferraro}
  \affil{Osservatorio Astonomico di Bologna, Via Ranzani 1, I-40127
Bologna, Italy} 


\begin{abstract}
We present wide field multiband photometry of the very low surface
brightness dwarf Spheroidal galaxy (dSph) Sextans. We have found
convincing evidence for the presence of multiple stellar populations
in this galaxy. In particular we discovered: {\em (a)} a Blue
Horizontal Branch tail kinked over the instability strip; {\em (b)} a
clearly bimodal distribution in color of the Red Giant Branch (RGB)
stars and {\em (c)} a double RGB-bump.

All these features suggest that (at least) two components are present
in the old stellar population that dominates the galaxy: a main one
with [Fe/H]$\sim-$1.8 and a minor component around [Fe/H]$\sim-$2.5.
Similar evidence has been recently found by Majewski et al. (1999) for
the Sculptor dwarf Spheroidal, suggesting that multiple star formation
episodes are common also in the most nearby dSphs that ceased their
star formation activity at very early epochs. \footnote{Based on
observations made with the European Southern Observatory facilities at
La Silla, Chile, using the WFI as part of the observing program
62.L-0354; also based on ESO WFI data obtained from the ESO/ST-ECF
Science Archive Facility.}
\end{abstract}


%
\section{Metallicity Puzzle in Sextans}

According to Mateo et al. (1991) and to Da Costa et al. (1991),
Sextans is dominated by a very old population, with a red Horizontal
Branch (HB) and several Blue Straggler Stars (BSS). They found
[Fe/H]$\sim-$1.6/$-$1.7 for this population, in disagreement with the
correlation between [Fe/H] and {\em M}$_V$ followed by the other dwarf
spheroidals.

Suntzeff et al. (1993), Geysler \& Sarajedini (1996), Shetrone et
al. (2001), found a metallicity of [Fe/H]$\sim-$2.0/$-$1.6, with an
intrinsic spread of $\sim$0.19 {\em dex}. The mean [Fe/H]$\sim-$2.0 is
in good agreement with the [Fe/H] vs. {\em M}$_V$ relation.

A variable stars survey (Mateo et al. 1995) confirm the presence of a
very old dominant population, but argue that the BSSs trace a younger
population, comprising $\sim$25$\%$ of the whole stellar content.

\begin{figure}
\plottwo{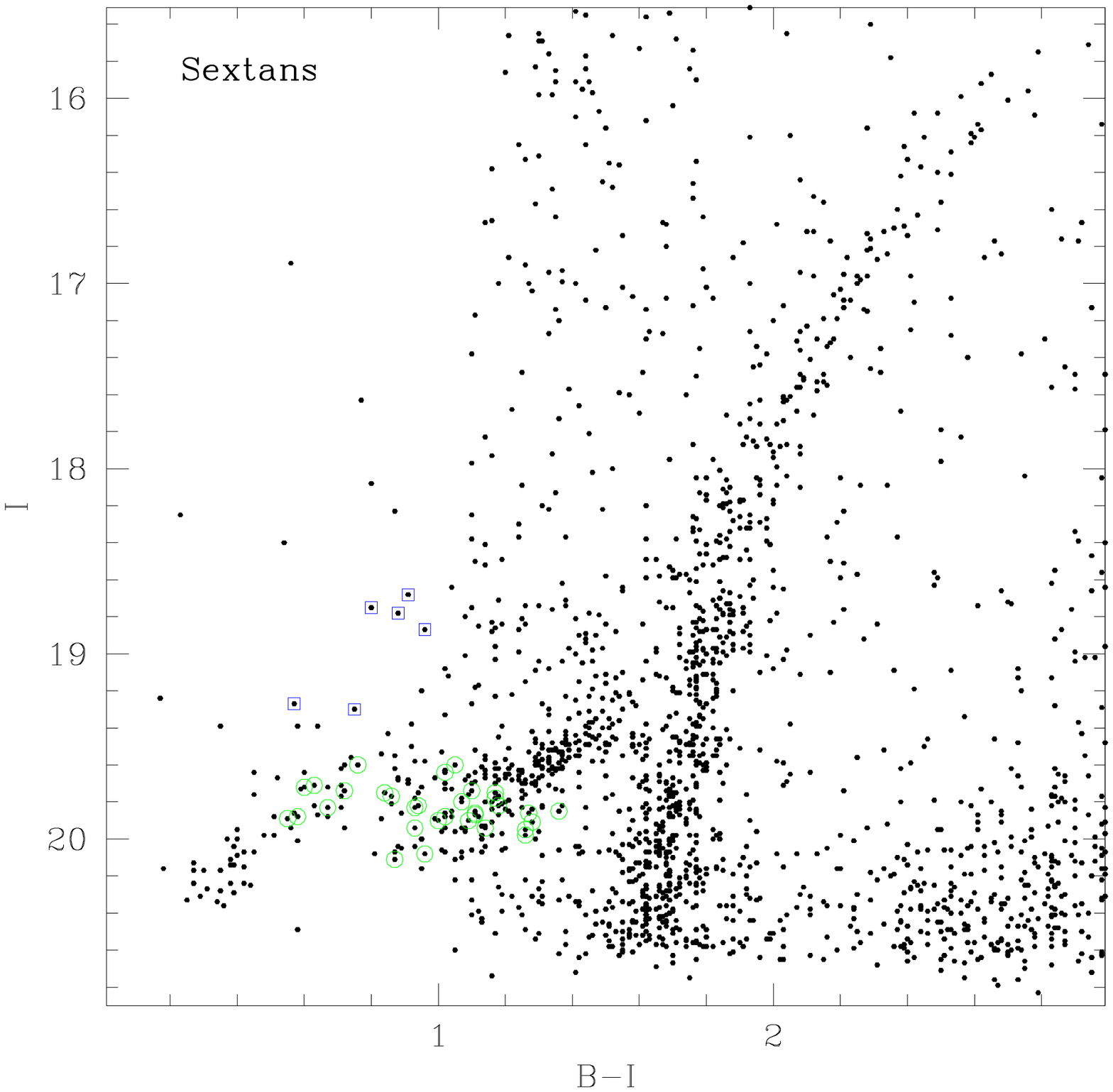}{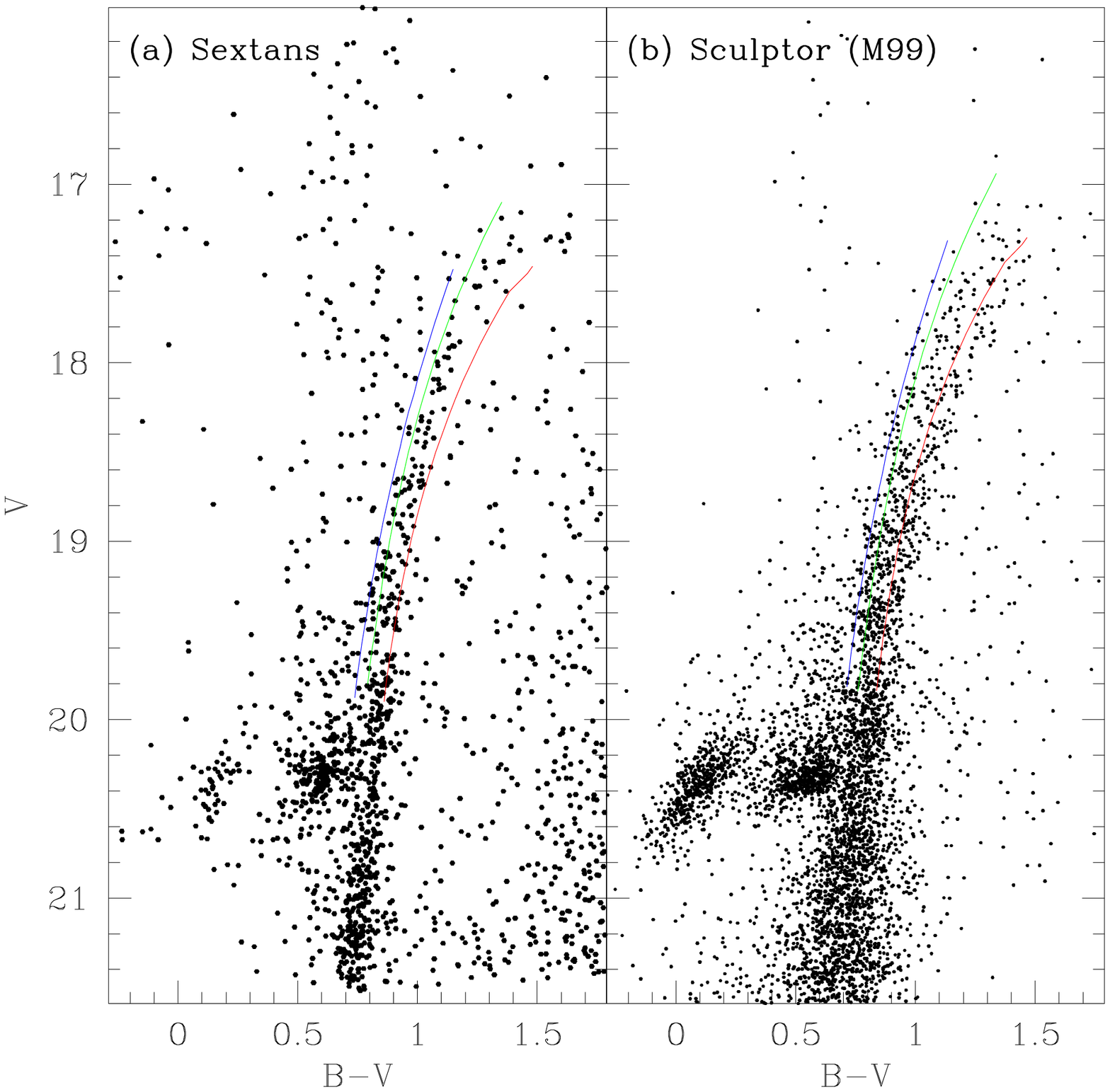}
\caption{{\em Left:} $(B,B-I)$ CMD of Sextans. Variable stars from
Mateo et al. (1995) are marked: RR Lyrae are enclosed by empty circles
and anomalous Cepheids by empty squares. {\em Right:} Our $(B,B-V)$
CMD is compared with the one of Sculptor from Majewski et
al. (1999). The mean ridge lines of NGC~5053, M~68 and M~3 from
Ferraro et al. (1999) are overplotted for reference (see text).}
\end{figure}

\section{Observations and Data Reduction}

All data have been taken with the Wide Field Imager (WFI) at the 2.2
{\em m} ESO-MPI Telescope, a mosaic CCD camera with a total field of
view of $\sim$34$'\times$33$'$.

Two 600 {\em sec} exposures have been obtained in the {\em B} and {\em
I} bands in January 1999. The observing run, observational set-up and
photometric calibration are the same as in Pancino et al. (2000).

One 300 {\em sec} exposure in the {\em V} band has been retrieved from
the WFI archive. The absolute calibration of the {\em V} band was
obtained by using the stars in common with Mateo et al. (1995).

All data have been processed with IRAF\footnote{IRAF is distributed by
the National Optical Astonomy Observatories, which is operated by the
association of Universities for Research in Astronomy, Inc., under
contract with the National Science Foundation.} and the photometry has
been performed with DoPHOT (Schechter, Mateo \& Saha 1993).

\begin{figure}
\plottwo{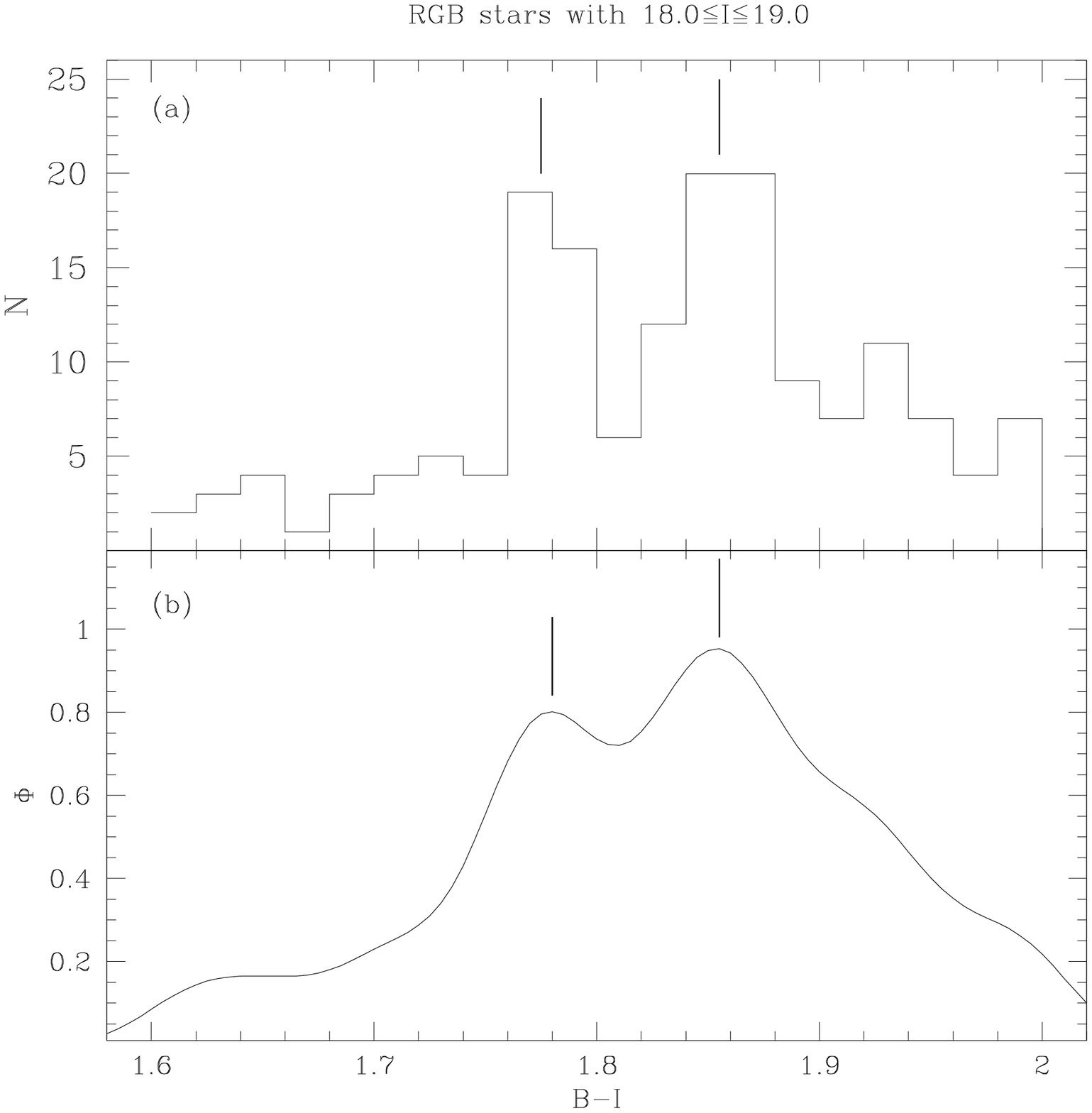}{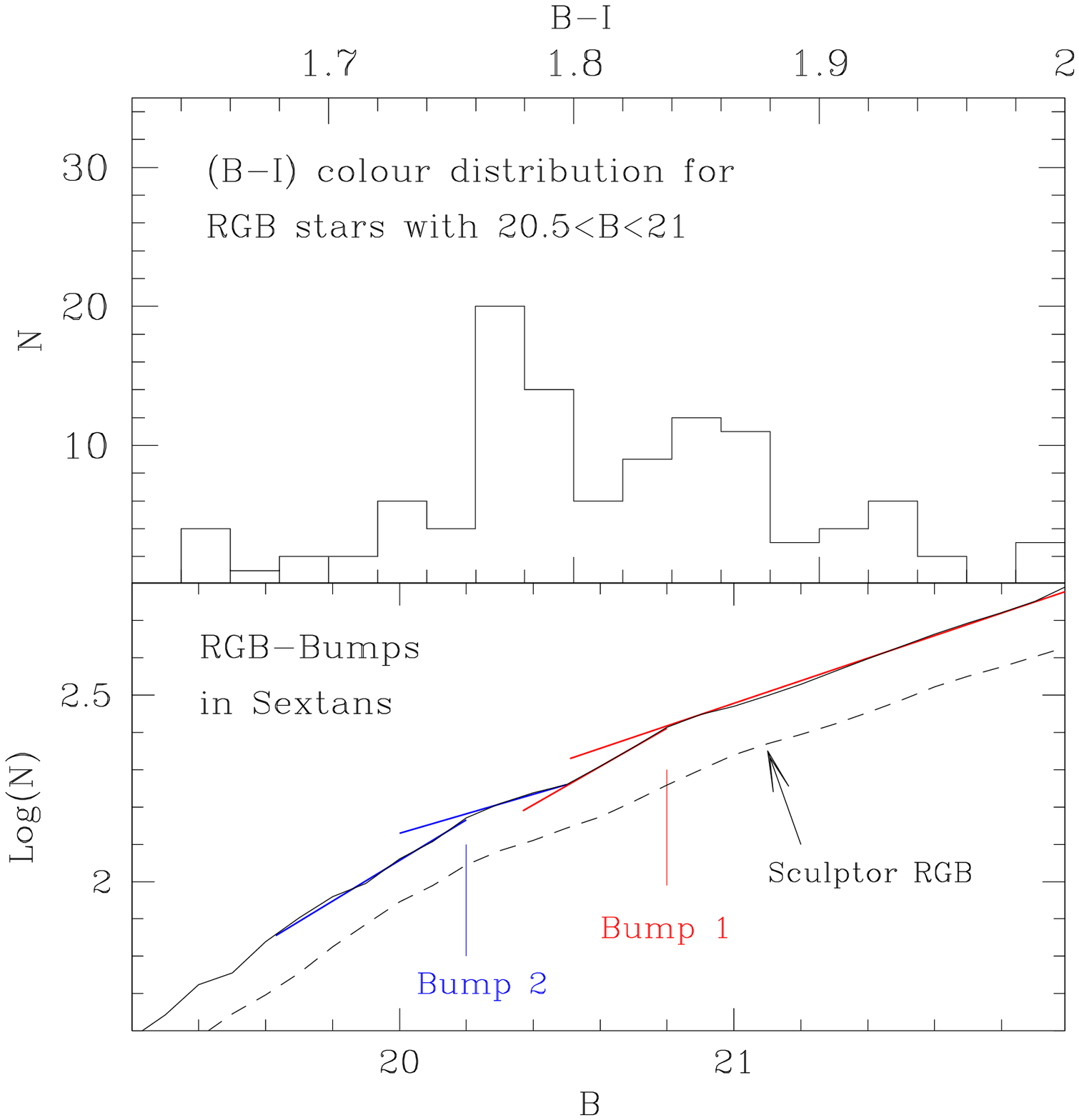}
\caption{{\em Left:} The color distribution of RGB stars with
18.0$\leq$I$\leq$19.0 is shown as a normal (top panel) and as a
generalized (bottom panel) histogram. The two main peaks are
marked. {\em Right:} The color distribution for stars with
20.5$\leq$B$\leq$21 confirms the RGB bimodality in color (top
panel). The cumulative luminosity function of Sextans is shown in the
bottom panel as a solid line, while the one of Sculptor is shown as a dashed
line. Two distinct RGB bumps are identified (see text).}
\end{figure}

\section{The Blue and Kinked HB}

As shown in Figure~1 above, an extended Blue HB can be observed for the
first time. The Blue HB is kinked over the instability strip, strongly
resembling the color-magnitude diagram observed for Sculptor by
Majewski et al. (1999).

They interpreted the Sculptor HB morphology as the signature of the
presence of two distinct stellar populations, one with
[Fe/H]$\sim-$2.0 and the other one with [Fe/H]$\sim-$1.5.

The RGB ridge lines of three template globular clusters from Ferraro
et al. (1999) have been overplotted on both the Sextans and the
Sculptor CMD in Figure~1 ({\em Right Panels}). From the bluest to
the reddest they are:
\begin{itemize}
\item{NGC~5053, [Fe/H]$\sim-$2.51}
\item{NGC~4590 (M68), [Fe/H]$\sim-$1.99}
\item{NGC~5727 (M3), [Fe/H]$\sim-$1.34} 
\end{itemize}

From this simple comparison, we can notice that the RGB is broad, that
the majority of stars lies on the blue side of the M3 ridge line and
that only $\sim$25$\%$ of the stars lie between the mean ridge lines
of M68 and NGC~5053.

\section{The Bimodal RGB Distribution}

Figure~1 suggests us that two separate populations with different
metallicities can be present in the Sextans dwarf Spheroidal
galaxy. Let us investigate further.

If we take now the most ``vertical'' section of the RGB and produce a
histogram of the color distribution, we see a clear bimodality (see
Figure~2, {\em Left Panels} and {\em Right Top Panel}).

From all of the above evidence, we can conclude that:
\begin{itemize}
\item{[Fe/H]$\simeq-$1.6 is at most an upper limit for the stars in Sextans}
\item{The majority of stars in Sextans has a mean [Fe/H]$\sim-$1.8} 
\item{Only $\sim$25$\%$ of the stars in Sextans seem to have
a lower [Fe/H]$<-$2.1. This fraction is compatible with the fraction
of stars in the blue HB ($\sim$17$\%$)}
\end{itemize}

\section{The Double RGB Bump}

As a last piece of evidence, we plotted the luminosity function for
the Sextans RGB (Figure~2, {\em Right Bottom Panel}), following the
standard approach as in Fusi Pecci et al. (1990) and we identified a
double RGB bump.

The luminosity of the RGB bump is known to depend primarily on the
metal content, and only weekly on age: more metal deficient stellar
populations have brighter bumps. Hence the double bump can be
interpreted as another signature of two different populations with
different metallicity.

Assuming (m$-$M)$_V=$19.79$\pm$0.15 from Mateo et al. (1995), we
derive M$_V^{B1}=$0.16 and M$_V^{B2}=-$0.44 for the faint and bright
bump respectively. Using
\begin{center}
M$_V=$0.87[Fe/H]$_{ZW}+$1.74
\end{center}
from Ferraro et al. (1999), we obtain [Fe/H]$_{ZW}^{B1}\sim-$1.8 and
[Fe/H]$_{ZW}^{B2}\sim-$2.5 for the two associated populations.


\end{document}